\documentclass[%
 reprint,
 superscriptaddress,
nofootinbib,
 floatfix,
 amsmath, amssymb,
 aps,
 prx,
]{revtex4-2}

\usepackage{ragged2e}
\usepackage{enumerate}
\usepackage{graphicx}
\usepackage{caption}
\usepackage{subcaption}
\usepackage{dcolumn}
\usepackage{bm}
\usepackage[colorlinks=true,linkcolor=blue,citecolor=blue,urlcolor=blue]{hyperref}
\usepackage{amsthm}
\newtheorem*{main_res}{Main result}
\usepackage{amsmath}
\usepackage{amssymb}
\usepackage{soul}
\usepackage{dsfont}
\usepackage{comment}
\usepackage{xcolor}
\usepackage{amsmath}
\usepackage{xcolor}
\definecolor{BlueGreen}{rgb}{0.0, 0.87, 0.87}
\usepackage{braket}
\usepackage{mathtools}
\usepackage[linesnumbered,ruled,vlined]{algorithm2e}

\theoremstyle{plain}

\theoremstyle{plain}

\theoremstyle{plain}
\newtheorem*{lem*}{\protect\lemmaname}
\theoremstyle{plain}

\theoremstyle{plain}

\usepackage[USenglish]{babel}
\providecommand{\corollaryname}{Corollary}
\providecommand{\lemmaname}{Lemma}
\providecommand{\propositionname}{Proposition}
\providecommand{\remarkname}{Remark}
\providecommand{\theoremname}{Theorem}

\usepackage[normalem]{ulem}

\definecolor{purple_palette}{HTML}{3b1c6e}
\hypersetup{colorlinks=true, 
            linkcolor=purple_palette,
            citecolor=purple_palette,
            urlcolor=purple_palette
}
\begin{document}

\preprint{APS/123-QED}

\title{
        Hamiltonian learning via quantum Zeno effect
        }

\author{Giacomo Franceschetto$^{\ast}$}

\affiliation{ICFO-Institut de Ciencies Fotoniques, The Barcelona Institute of Science and Technology, 08860 Castelldefels (Barcelona), Spain}

\author{Egle Pagliaro$^{\ast}$}

\affiliation{ICFO-Institut de Ciencies Fotoniques, The Barcelona Institute of Science and Technology, 08860 Castelldefels (Barcelona), Spain}

\author{Luciano Pereira}
\affiliation{ICFO-Institut de Ciencies Fotoniques, The Barcelona Institute of Science and Technology, 08860 Castelldefels (Barcelona), Spain}

\author{Leonardo Zambrano}
\affiliation{ICFO-Institut de Ciencies Fotoniques, The Barcelona Institute of Science and Technology, 08860 Castelldefels (Barcelona), Spain}
 
\author{Antonio Acín}
\affiliation{ICFO-Institut de Ciencies Fotoniques, The Barcelona Institute of Science and Technology, 08860 Castelldefels (Barcelona), Spain}
\affiliation{ICREA, Passeig Lluis Companys 23, 08010 Barcelona, Spain}

\begingroup
\renewcommand{\thefootnote}{$\ast$}
\footnotetext{These two authors contributed equally}
\endgroup
\begingroup
\renewcommand{\thefootnote}{$$}
\footnotetext{Authors' contacts: \texttt{name.surname@icfo.eu}}

\date{\today}
\begin{abstract}
Determining the Hamiltonian of a quantum system is essential for understanding its dynamics and validating its behavior. Hamiltonian learning provides a data-driven approach to reconstruct the generator of the dynamics from measurements on the evolved system. Among its applications, it is particularly important for benchmarking and characterizing quantum hardware, such as quantum computers and simulators. However, as these devices grow in size and complexity, this task becomes increasingly challenging. To address this, we propose a scalable and experimentally friendly Hamiltonian learning protocol for Hamiltonian operators made of local interactions. It leverages the quantum Zeno effect as a reshaping tool to localize the system's dynamics and then applies quantum process tomography to learn the coefficients of a local subset of the Hamiltonian acting on selected qubits. Unlike existing approaches, our method does not require complex state preparations and uses experimentally accessible, coherence-preserving operations.
We derive theoretical performance guarantees and demonstrate the feasibility of our protocol both with numerical simulations and through an experimental implementation on IBM’s superconducting quantum hardware, successfully learning the coefficients of a 109-qubit Hamiltonian.

\end{abstract}

\maketitle

\section{Introduction}

A central challenge in quantum science is to understand and characterize the dynamics of closed physical systems, which are entirely governed by their Hamiltonians. In realistic settings, however, the effective Hamiltonian is often unknown and must be inferred from experimental observations, a problem addressed by Hamiltonian learning (HL) \cite{Wiebe_2014_HL}. This task is critical not only for uncovering the fundamental structure of many-body quantum systems, but also for a broad range of applications including device certification, error diagnosis, and quantum control optimization \cite{rouze2024learning, Bairey_2019_localmeasurements, Haah_2024_gibbstates, yu2023robust}. One important application of HL is the benchmarking of quantum simulators \cite{Pastori2022}, engineered quantum systems designed to mimic complex quantum dynamics and probe regimes beyond the reach of classical computation \cite{Georgescu_2014_quantumsimulators, Altman_2021_quantumsimulators}. As these systems grow in size and sophistication, with platforms now reaching over a hundred qubits, accurately identifying the effective Hamiltonians they realize becomes essential for validating their performance and establishing trust in their predictive power \cite{Eisert_2020_cetification}. This growing complexity requires efficient Hamiltonian learning techniques that remain experimentally doable, even at scale.

However, existing HL protocols typically face significant practical limitations. Many rely on the ability to prepare specific quantum states, such as eigenstates or thermal (Gibbs) states of the unknown Hamiltonian \cite{Burgarth_2009_restrictedaccess, Anshu_2021_gibbs, Haah_2024_gibbstates, Bairey_2019_localmeasurements, Qi_2019_fromeigenstate}. Preparing such states is experimentally demanding, particularly on noisy intermediate-scale quantum devices, limiting the applicability of these approaches.

As a more practical alternative, black-box unitary models treat the system as an unknown process accessible only through simple initial state preparation, black-box application of the Hamiltonian dynamics, and measurements \cite{Gu_2024_blackboxHL}. While these approaches avoid complex state preparation, the learning task remains fundamentally challenging due to the exponential growth of the Hilbert space with system size. 

A promising direction is to leverage the intrinsic structure of many physical systems. In particular, a large class of relevant models is governed by geometrically local Hamiltonians, where interactions are limited to nearby qubits on a spatial layout. This structure provides a natural advantage: focusing on local interactions reduces the complexity of the reconstruction task \cite{dasilva2011}. Following this direction, recent works have introduced Hamiltonian reshaping strategies, by which we mean procedures that alter the system’s dynamics in a controlled way, suppressing unwanted couplings while preserving an effective Hamiltonian whose parameters can still be traced back to those of the original model. A standard method applies random Pauli operations during the evolution to average out undesired terms \cite{Huang_2023_heisemberscaling}, akin to dynamical decoupling techniques \cite{Wang_2015_dynamicaldecoupling}. However, these protocols often require demanding operations, such as frequent gate applications or fine-grained control over specific Hamiltonian components, which are challenging to implement on current hardware. Moreover, such control sequences typically increase the total evolution time, pushing the system closer to decoherence.

In this work, we present a Hamiltonian learning protocol specifically tailored to geometrically local systems and designed to mitigate decoherence overhead. The method enables scalable and parallel characterization of local Hamiltonian terms by exploiting the quantum Zeno effect (QZE), which refers to the effective freezing of a quantum system’s evolution through frequent interventions \cite{Facchi_2002}. While originally associated with repeated projective \cite{Kakuyanagi_2015_projmeas,Kalb2026_projmeas} or continuous \cite{Slichter_2016_contmeas} measurements , the QZE can also be realized through frequent unitary kicks \cite{Facchi_2004, Facchi_2008_unitarykicks}. This unitary version avoids destructive measurements and is particularly suitable for quantum computing platforms, where such kicks can be efficiently implemented using virtual $Z$ gates \cite{McKay_2017_virtualZgates,Vezvaee2025_virtual}. Our protocol exploits QZE with unitary kicks to dynamically isolate local interactions in the target Hamiltonian. By applying frequent virtual $Z$ gates to selected qubits, we suppress unwanted Hamiltonian interactions and reshape the dynamics, effectively decoupling the system into non-interacting $n$-qubit patches. Each patch is then independently characterized using quantum process tomography (QPT), allowing for accurate reconstruction of the corresponding local Hamiltonian terms. Moreover, since non-overlapping patches evolve independently, they can be characterized in parallel \cite{Cotler_2020_QOT, Ara_jo_2022_localQOT, Pereira2023_parallelQT}, enabling efficient scaling with system size (Figure \ref{fig:intro}). The protocol is experimentally friendly, requiring only product states preparation, short-time evolution, and local Pauli measurements, with the additional advantage of using virtual $Z$ gates, which execute instantaneously and do not introduce decoherence.
We derive theoretical performance bounds on the protocol's performance, simulate it on systems up to 128 qubits, and demonstrate its feasibility through an experimental implementation on IBM’s 127-qubit superconducting quantum processor.

The article is organized as follows. Section~\ref{sec: preliminaries} introduces the Hamiltonian learning problem and presents our method’s core components: the quantum Zeno effect via unitary kicks and quantum process tomography. A complete description of the protocol, focusing on geometrically two-qubit local Hamiltonians, is given in Section~\ref{sec:protocol}. Section~\ref{sec:bound} derives analytical bounds on the performance; Section~\ref{sec:numerics} presents numerical simulations that validate the protocol’s accuracy and robustness; and Section~\ref{sec:experimental} details the experimental implementation on IBM hardware. Finally, Section~\ref{sec:discussion} concludes with a discussion of the results and potential future directions.

\begin{figure}[h!t]
    \includegraphics[width=0.42\textwidth]{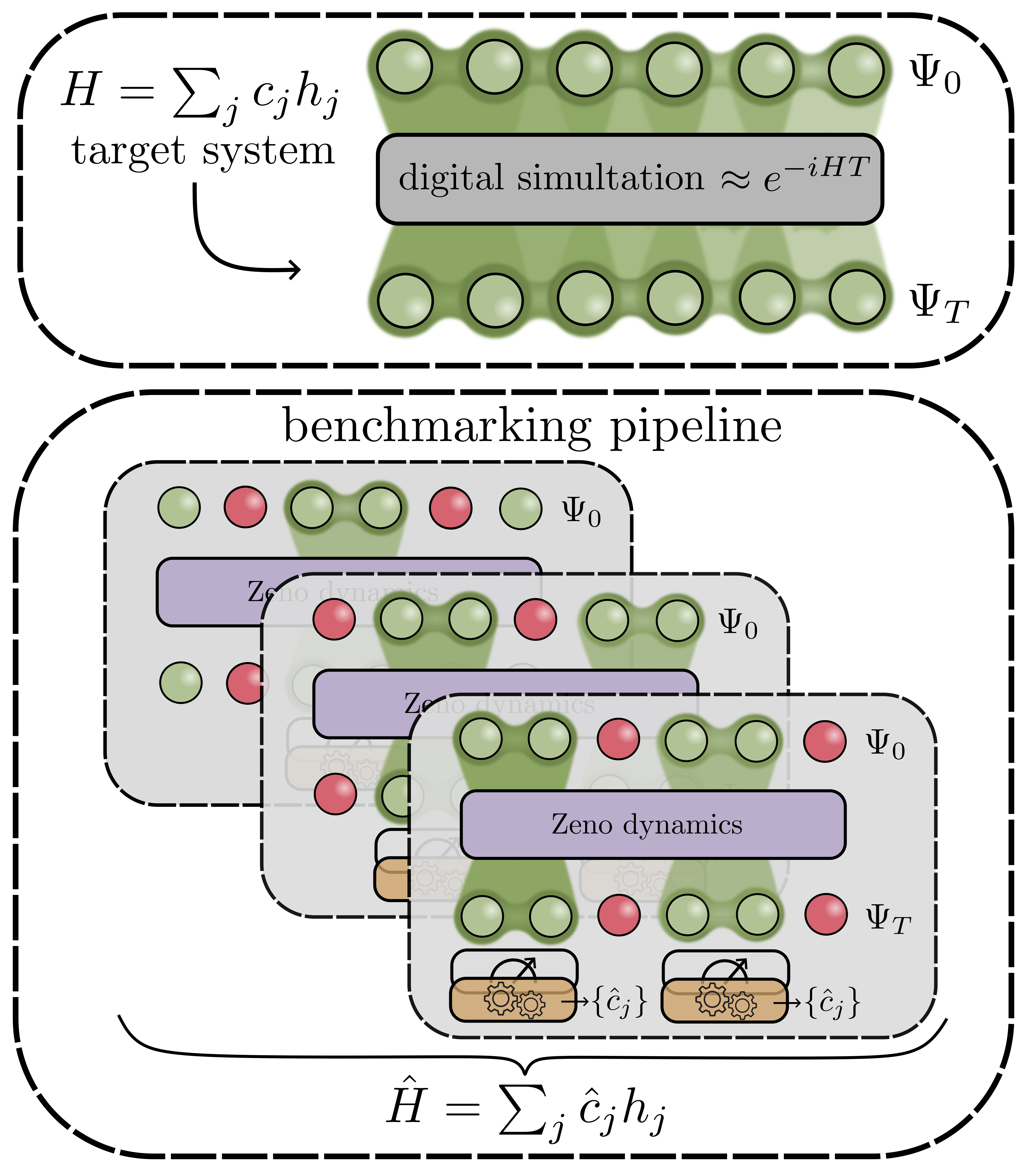}
    \caption{\justifying Benchmarking digital quantum simulators via Hamiltonian learning. Top: a digital quantum simulator implements the target evolution. Bottom: our protocol uses Zeno-based reshaping to isolate local patches, enabling parallel quantum process tomography and accurate reconstruction of the simulated Hamiltonian.} \label{fig:intro}
\end{figure}

\section{Preliminaries}
\label{sec: preliminaries}

We begin by stating the learning problem addressed in this work. Our goal is to estimate the coefficients of a quantum many-body Hamiltonian, given access to time-evolution unitaries generated by it. More precisely, we assume that the Hamiltonian can be expressed in a known operator basis as
\begin{equation}\label{eq:hamiltonian}
H = \sum_{j} c_j h_j,
\end{equation}
where each $h_j$ is a known local operator acting nontrivially on at most $k$ qubits within a fixed interaction range determined by the system’s geometry, i.e., the Hamiltonian is geometrically $k$-local. The task is to learn the unknown real coefficients $c_j$
that fully characterize $H$.
We outline two technical tools central to our protocol, and their associated precision bounds. First, we describe the QZE, where frequent unitary kicks induce an effective confined evolution. Next, we introduce QPT as a method to reconstruct an unknown quantum process. Each of these tools comes with its own resource scaling, determining the requirements to reach a target precision in its respective application.
\subsection{Quantum Zeno effect via unitary kicks}
\label{sub:ham_reshaping}
The QZE describes how frequent interventions can inhibit or modify a system's natural evolution, effectively constraining it to a restricted subspace of the Hilbert space. 
While it was originally formulated in terms of repeated projective measurements, the same mechanism can be realized in a fully unitary way, for example through strong continuous couplings or via instantaneous unitary kicks. 
In our protocol, we adopt the unitary-kick implementation, which offers two key advantages: it preserves unitarity throughout the process and it is well-suited to hardware platforms. Concretely, over a total evolution time \(T\), we interleave the system's natural dynamics under the unknown Hamiltonian \(H\) with a sequence of \(r\) instantaneous kicks \(U_{\text{kick}}\), applied at intervals \(T/r\). At the end of the sequence, we apply the inverse operation $ \left (U_{\text{kick}}^\dagger \right ) ^r$ (the ``back-kick'') to remove the trivial phases accumulated from the repeated kicks, so that only the effective Zeno evolution remains.
Formally, we define the kicked evolution as
\begin{align}
    V_r(T) = \left (U_{\text{kick}}^\dag \right )^r \left (U_{\text{kick}}\, e^{-i H T / r}\right)^r.
\end{align}
It can be proven that, in the limit $r \rightarrow \infty$, this sequence converges to
\begin{align}
    U_{\text{Z}}(T) = e^{-i H_{\text{Z}} T},
\end{align}
where the effective Hamiltonian $H_{\text{Z}} = \sum_k P_kHP_k$
is the so-called Zeno Hamiltonian, and $P_k$ are the spectral projectors of $U_{\text{kick}}$.
Thus, in this limit, the dynamics is effectively confined to the eigenspaces of the kick operator.

The convergence of the kicked evolution to the effective dynamics generated by $H_{\text{Z}}$ can be rigorously bounded. Reference \cite{Burgarth_2022} establishes that
\begin{equation}
    \|V_r(T) - U_Z(T)\|_{\rm op} \;\le\; \frac{C_{\text{Z}}}{r},
\end{equation}
with
\begin{equation}
    C_{\text{Z}} \;=\;2\left(\frac{\sqrt{m}}{\xi}+1\right) T\|H\| (1+2T\|H\|),
\end{equation}
and the norm $\|\cdot\|_{\mathrm{op}}$ is taken as the spectral norm. Here, $m$ denotes the number of Zeno subspaces determined by the spectral projectors of $U_{\text{kick}}$, and $\xi$ quantifies its inverse spectral gap.
This bound implies that, to reach accuracy $\epsilon_{\text{Z}}$, it suffices to choose
\begin{equation}
    r \;\ge\; \frac{C_{\text{Z}}}{\epsilon_{\text{Z}}}.
\end{equation}

This result quantifies how well the kicked evolution approximates the target dynamics after projecting onto Zeno subspaces. The design of the kick operator determines the structure of the Zeno subspaces that confine the effective dynamics, allowing us to selectively suppress or preserve terms of the original Hamiltonian and thereby control the system’s effective evolution. This control has made the QZE a valuable tool across quantum technologies, with applications ranging from decoherence suppression \cite{Facchi_2004} and subspace stabilization \cite{Facchi_2002} to error correction \cite{Erez_2004}. 
\begin{figure*}[t!]
    \centering
    \includegraphics[width=\linewidth]{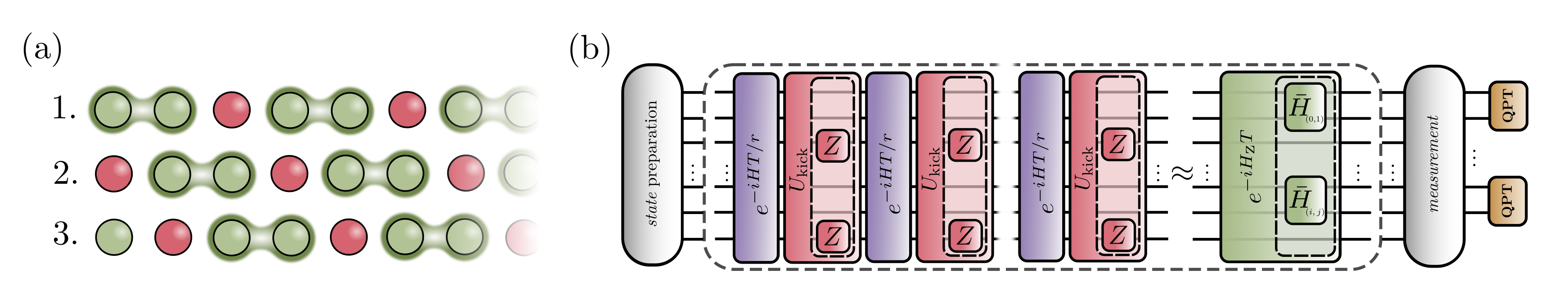}
    \caption{\justifying Sketch of the protocol. (a) Three reshaping configurations are required to learn an $ N$-qubit Hamiltonian with $k=2$ and a linear geometry. Red circles represent the frozen qubits, while the green blurred regions represent the learned interactions. (b) Circuit-like scheme of the protocol for the first reshaping configuration. The effect of alternative application of unitary evolution and unitary kicks can be approximated by the evolution under the Zeno Hamiltonian, where different pairs of target qubits evolve independently. By applying this procedure to a set of initial states and performing a selected set of measurements at the end, the learning of the interactions in each target subsystem can be achieved. Note that no back-kick is shown in the scheme, since in our setting ($Z$ kicks with even $r$) the accumulated kick factor is the identity.}
    \label{fig:protocol_wide}
\end{figure*}

\subsection{Quantum process tomography}
\label{sub:coeff_learning}
QPT is a standard technique for reconstructing an unknown quantum channel by probing it with carefully chosen input states, known as fiducial states, and measuring the corresponding outputs \cite{Chuang_1997, Poyatos_1997, Childs_2001}.
Formally, for a channel $\varLambda$ acting on an $n$-qubit patch over a total time $T$, QPT reconstructs $\varLambda$ by preparing a set of linearly independent fiducial states ${\rho_j}$, applying the channel to each, and measuring an informationally complete collection of positive operator-valued measures (POVMs) ${E_k}$. Under these conditions, the outcome probability matrix $P$ can be expressed as
\begin{align}
    P &= \sum_{k,j} \langle\!\bra{E_k}\,\varLambda(\rho_j)\,\rangle\!\rangle\ket{k}\bra{j} \\
    &= \biggl(  \sum_{k} \ket{k}\langle\!\bra{E_k}\,\biggr)\Upsilon\biggl(\sum_{j} |\rho_j\,\rangle\!\rangle\bra{j}\biggr) \\
    &= M\,\Upsilon\,S,
\end{align}
where $\Upsilon$ is the Choi matrix of $\varLambda$ and double kets denote vectorization. Since $M$ and $S$ are known by construction and an approximation of $P$ is obtained from experiment, one can estimate $\Upsilon$ by linear inversion:
\begin{equation}
    \hat{\Upsilon} = M^{-1}\,\hat{P}\,S^{-1}.
\end{equation}

Furthermore, combining the results from \cite{surawy2022projected, oufkir_sample-optimal_2023}, one can bound the sample complexity of QPT. Specifically, for $\epsilon_\text{QPT}, \delta \in (0, 1)$,  
\begin{equation}
   \operatorname{Pr} \left (|| \hat{\varLambda}-\varLambda||_\diamond
   \le \epsilon_\text{QPT} \right )\ge 1-\delta,
\end{equation}
provided the total number of copies satisfies
\begin{equation}
    N_{\text{copies}} \geq \frac{C_\text{QPT}(n)}{\epsilon_\text{QPT}^2}\ln\left( \frac{2^{4n}}{\delta} \right),
\end{equation}
where
\begin{equation}
    C_\text{QPT} (n) = \frac{8}{3} 3^{2n}\, 2^{4n}.
\end{equation}
In our protocol, this procedure is applied after the dynamics have been reshaped via the QZE. The chosen Zeno dynamics, engineered via the kick operator, partitions the system into spatially disjoint $n$-qubit patches, on which QPT can be performed independently and in parallel on each patch, thereby enabling an efficient reconstruction of the local evolutions.

\section{Protocol overview}
\label{sec:protocol}

Building on the preliminaries outlined in Section \ref{sec: preliminaries}, here we detail the implementation of our HL protocol. 
Our procedure combines the QZE as a Hamiltonian reshaping tool (\ref{sub:ham_reshaping}) with QPT as a reconstruction routine (Sec.~\ref{sub:coeff_learning}) to efficiently recover geometrically $k$-local Hamiltonians. For clarity, we illustrate the method in the case of geometrically 2-local Hamiltonians on a one-dimensional (1D) chain, though the approach extends to larger $k$ and higher-dimensional geometries. A schematic overview of the protocol is provided in Fig.~\ref{fig:protocol_wide}. 

The protocol consists of three stages: (1) preprocessing, where the reshaping configurations are defined according to the geometry and locality of the Hamiltonian; (2) processing, where the Zeno dynamics corresponding to each configuration are implemented while collecting the data required for quantum process tomography; and (3) postprocessing, where the measurement data are used to reconstruct the coefficients for each configuration and combine them to obtain the complete Hamiltonian.

In the preprocessing stage, we define the reshaping configurations required to isolate the desired interactions. These configurations are chosen so that every Hamiltonian term is not suppressed in at least one configuration, ensuring that all coefficients can be independently learned and later combined to reconstruct the full Hamiltonian. For each configuration, we must specify which qubits serve as target patches (whose interactions we aim to learn) and which are subject to the unitary kicks that induce the global Zeno effect, effectively decoupling them from the rest of the qubits. For $k=2$, this means fixing three distinct configurations, displayed in Fig.\thinspace\ref{fig:protocol_wide}(a). The first configuration assigns the first two qubits as target patches and designates the next qubit as the decoupling qubit. We then apply the same assignment for the subsequent qubits in the chain. The remaining two configurations are constructed similarly, beginning with the second and third qubits, respectively.  

In the processing stage, we implement the Zeno dynamics for each reshaping configuration and collect the measurement data required for quantum process tomography. The system is initialized in an $N$-qubit product state, where the qubits in the target region (highlighted in green in Fig.~\ref{fig:protocol_wide}.a) are prepared in arbitrary two-qubit product fiducial states $|\psi_z\rangle$, while all qubits outside the target region (shown in red) are initialized in the $Z$-eigenstate $|0\rangle$. For instance, in a chain of $N=6$ qubits, a possible configuration selects $(0,1)$ and $(3,4)$ as target regions, while the rest are fixed in $|0\rangle$:
\begin{align}
    |\Psi_0\rangle = |\psi_z\rangle_{0,1} \otimes |0\rangle \otimes |\psi_z\rangle_{3,4} \otimes |0\rangle 
\end{align}

The system then undergoes $r$ sequences of free evolution under the full Hamiltonian alternated with unitary kicks $U_{\text{kick}}$, which induce a QZE that dynamically decouples the target qubits from the rest of the system. For this configuration, to induce the intended dynamics, we construct the kick operator as $U_{\text{kick}} = \mathbb{I}_2 \otimes Z \otimes \mathbb{I}_2 \otimes Z$, acting on the non-target qubits. Repeated applications of this unitary freeze the dynamics of these qubits, while allowing the target patches to evolve under a local effective Hamiltonian. This occurs because the kicks suppress transitions between the eigenspaces of $U_{\text{kick}}$, effectively confining the evolution of the system to the target regions.

To clarify how the reshaping acts at the local level, consider a single non-target qubit $i$, and suppose the Hamiltonian is expressed in terms of single- and two-qubit Pauli operators ($\sigma_i$ and $\sigma_i\sigma_j$ for $\sigma \in \{X,Y,Z\}$). To decouple qubit $i$ from its neighbors, we apply the unitary kick $U_{\rm kick} = Z_i$. In the limit of frequent kicks, the effective Hamiltonian no longer contains two-qubit terms of the form $\sigma_j\sigma_i$ with $j$ adjacent to $i$ and $\sigma_i \neq Z$. Instead, two-qubit terms commuting with $Z_i$, which should ideally be removed to isolate the local dynamics, survive and are reshaped into additional contributions to the single-qubit coefficients of the neighboring qubits. All other terms—single-qubit operators on $i$ and interactions not involving $i$—remain unchanged. As a result, the dynamics factorize into independent blocks, enabling each patch to be characterized separately.

This reshaping strategy is particularly convenient in practice when the kicks are chosen as $Z$ rotations, as in the above example, since several hardware platforms support virtual $Z$ gates. Rather than applying a physical microwave pulse, the control electronics advance the global phase reference by $\pi$, which shifts all subsequent $X$ and $Y$ drives accordingly, and is mathematically identical to a true $Z(\pi)$ rotation. Because no microwave pulse is required and the phase register is maintained with high precision, virtual $Z$ gates execute effectively instantaneously and without added decoherence, making them highly efficient.

Once the target patches have been dynamically isolated via the Zeno procedure, measurements are performed on the states evolved under the effective dynamics. An informationally complete set of POVMs is applied independently and in parallel across the patches. The measurement outcomes are gathered to form the dataset used in the reconstruction.

In the postprocessing stage, this data is used to compute the Choi matrices $\hat{\Upsilon}$ for each patch via linear inversion as described in Section \ref{sub:coeff_learning}.
However, due to experimental noise, the reconstructed matrices do not, in general, correspond to valid physical quantum processes. Several methods have been proposed to project such noisy estimates onto the space of physical quantum channels \cite{Knee_QPS, surawy2022projected, Barbera_QPS, Quiroga2023_QPS, DiColandrea2024_QPS, Bao_QPS, Jaouni2024_QPS, Xiao_QPS}. In our case, we expect the true underlying process to be unitary and the noise to be small. Therefore, we assume $\varLambda$ to be unitary and obtain an estimate of the physical process by taking the rank-1 approximation of $\hat{\Upsilon}$, reshaping it into its superoperator form, and extracting its unitary polar factor $\hat{U}$. The effective Hamiltonian of the patch is then retrieved as
\begin{equation}
    \hat{H}_\text{patch} = -\frac{i}{T}\,\log(\hat{U}),
\end{equation}
and the coefficients of interest are computed by projecting onto the known operator basis as $\operatorname{Tr}\bigl(h_j\,\hat{H}_\text{patch}\bigr)$.

This procedure efficiently learns the interactions acting on the target qubits, though it omits those involving non-target qubits. As explained above, commuting interactions that survive the kicks appear as additional single-qubit contributions on neighboring qubits, thereby modifying the boundary terms: some entries of the local effective Hamiltonians differ from the original ones due to residual boundary effects. By repeating the reshaping procedure over different reshaping configurations and applying the above reconstruction to each, we can recover every local interaction and correct these distortions. For 1D systems, the number of required configurations scales linearly with the interaction length and is constant over the system size.  For the 2-local case considered here, three reshaping configurations suffice to fully reconstruct the Hamiltonian, as illustrated in Figure~\ref{fig:protocol_wide} (a). A pseudo-code summary of the complete procedure is provided in Algorithm~\ref{alg:HL_Zeno}.

\begin{algorithm}   
\caption{}\label{alg:HL_Zeno}
\KwIn{Number of qubits $N$, unknown unitary evolution $U(t)$, terms $\{h_j\}$, number of kicks $r$, total time $T$}
Define reshaping configurations based on $N$\;
\tcp{{\footnotesize e.g.$\{$(target(2), kick(1), target(2), kick(1)),}}
\tcp{{\footnotesize \hspace{7.75mm}(kick(1), target(2), kick(1)), target(2)),}}
\tcp{{\footnotesize \hspace{8.5mm}$\cdots\}$}}
\For{each reshaping configuration}{
  \For{$\psi_{\text{Z}}$ in $\{\psi_{\text{Z}}\}$ (tomographically complete)}{
      \For{$E_k$ in $\{E_k\}$ (informationally complete)}{
        Prepare initial state $\Psi_0$\;
        \tcp{{\footnotesize e.g.$\, \ket{\psi_{\text{Z}}}\otimes \ket{0} \otimes \ket{\psi_{\text{Z}}}\otimes \ket{0}$}} 
        Evolve $\Psi_0$ with proper Zeno dynamics\;
        \tcp{{\footnotesize like $H_{\text{Z}} = \bar{H}_{(0,1)}^{}\otimes\mathbb{I}\otimes \bar{H}_{(3, 4)}^{}\otimes\mathbb{I}$}}
        \For{$r$ times}{
          Apply $U(T/r)$\;
          Apply $U_{\text{kick}}$\;
          \tcp{{\footnotesize e.g. $U_{\text{kick}} = \mathbb{I}_2\otimes Z\otimes\mathbb{I}_2\otimes Z$}}
        }
        Estimate expectation values of $E_k$ on the different target qubits\;
      }
    }
  Reconstruct local effective Hamiltonians via QPT\;
  Learn local coefficients on target qubits\;
}
Combine local coefficients\;
\KwOut{Estimated coefficients $\{\hat{c}_j\}$}
\end{algorithm}

\section{Perfomance guarantees}
\label{sec:bound}

\newtheorem{theorem}{Theorem}
We now establish performance guarantees for our protocol. Our first goal is to bound the reconstruction error of the local quantum channel describing the time evolution of each dynamically isolated target patch. The following theorem provides sufficient conditions to control the diamond-norm distance between the true unitary channel induced by the physical evolution and its reconstructed approximation, with separate contributions from imperfect Zeno reshaping and finite-sample noise in QPT.
\begin{theorem}\label{thm:1}
Let $\varLambda$ be the target unitary channel on an $n$-qubit subsystem. Suppose the global system evolves under
\[
    U = e^{-i H T},
\]
and that, by applying repeated unitary kicks on neighboring qubits, the target $n$-qubit region is isolated through the QZE. Denote by $\hat{\varLambda}_{\mathrm{LI}}$ the $n$-qubit channel obtained from linear inversion in QPT on the target subsystem, and by $\hat{\varLambda}_{\mathrm{U1}}$ its rank-1 projection onto the unitary set via the unitary polar factor.

Fix any $\epsilon,\delta\in(0,1)$ and split
\[
    \epsilon = \epsilon_{\text{Z}} + \epsilon_{\text{QPT}}.
\]
Then, to ensure
\[
    \Pr\Bigl[\|\hat{\varLambda}_{\mathrm{U1}} - \varLambda\|_\diamond 
    \;\le\; \|\hat{\varLambda}_{\mathrm{U1}} - \hat{\varLambda}_{\mathrm{LI}}\|_\diamond + \epsilon \Bigr]
    \;\ge\; 1 - \delta,
\]
\begin{figure*}[t]
    \centering
    \includegraphics[width=0.9\textwidth]{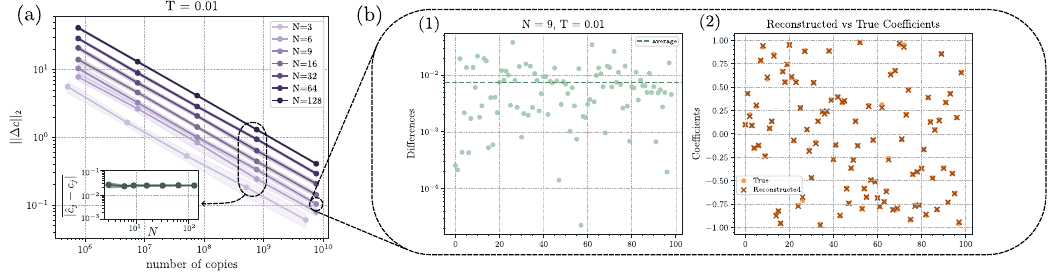}
    \caption{\justifying Numerical results (a). Euclidean norm of the difference between reconstructed and true Hamiltonian coefficients as a function of the total number of copies used in the tomography for system sizes up to  $N=128$ . The shaded areas indicate one standard deviation over 10 different Hamiltonians. We fix the evolution time to $T=0.01$ and number of kicks to $r=10$. In the bottom left insert panel, we show the average absolute error for the coefficients at a fixed number of copies $\approx10^{9}$
    (b). Absolute errors (1) and comparison of true versus reconstructed Hamiltonian coefficients (2) for $N=9$ and $\approx10^{10}$ total copies.} \label{fig:wide_plot}
\end{figure*}

it suffices to satisfy both:
\begin{enumerate}
    \item A total number of copies of
    \[
        N_{\mathrm{copies}} \;\ge\; \frac{C_{\mathrm{QPT}}(n)}{\epsilon_{\mathrm{QPT}}^2}\,\ln\!\Bigl(\frac{2^{4n}}{\delta}\Bigr),
    \]
    where \(C_{\mathrm{QPT}}\) is defined as in Section \ref{sub:coeff_learning}.
    \item A number of Zeno-kick repetitions
    \[
        r \;\ge\; \,\frac{2C_{Z}}{\epsilon_{\rm{Z}}},
    \]
    where \(C_{Z}\) is defined as in Section \ref{sub:ham_reshaping}.
\end{enumerate}
\end{theorem}
The term \(\|\hat{\varLambda}_{\mathrm{U1}} - \hat{\varLambda}_{\mathrm{LI}}\|_\diamond\), which quantifies the additional error due to the projection onto the set of unitary channels, can be evaluated a posteriori based on the tomography data and the unitary approximation step. Starting from this channel-level bound, we now derive a bound on the Hamiltonian coefficients, which is the primary quantity of interest for Hamiltonian learning.
\newtheorem{corollary}{Corollary}[theorem]
\begin{corollary}\label{coeff_bound}
Under the assumptions of the preceding theorem, and further assuming 
\[
    \|\bar{H}_\text{patch}\|_{\mathrm{op}},\;\|\hat{\bar{H}}_\text{patch}\|_{\mathrm{op}} \le \frac{1}{\pi T},
\]
write 
\[
    \bar{H}_\text{patch} = \sum_j c_j h_j,\quad \hat{\bar{H}}_\text{patch} = \sum_j \hat{c}_j h_j,\quad \Delta c = c - \hat{c}.
\]
Then, with probability at least \(1 - \delta\),
\[
    \|\Delta c\|_2 \;\le\; \frac{\pi}{T}\,\Bigl(\|\hat{\varLambda}_{\mathrm{U1}} - \hat{\varLambda}_{\mathrm{LI}}\|_\diamond + \epsilon\Bigr).
\]
\end{corollary}

The proofs of Theorem \ref{thm:1} and Corollary \ref{coeff_bound} can be found in Appendix \ref{sec::theoremproof} and \ref{sec::corproof}. 
Finally, the global Hamiltonian coefficients are obtained by combining the local reconstructions, correcting for reshaping side effects via additive or subtractive combinations. 
Let $n_p$ denote the total number of patches and define, for each patch $i$, the achieved precision on the local reconstruction
\begin{equation}
\varepsilon_i \;=\; \frac{\pi}{T}\Bigl(\,\|\hat{\varLambda}_{\mathrm{U1}}
- \hat{\varLambda}_{\mathrm{LI}}\|_\diamond + \epsilon \Bigr).
\end{equation}
Under the assumptions of Theorem~\ref{thm:1} and Corollary~\ref{coeff_bound}, we obtain the following guarantee on the overall reconstruction accuracy.

\begin{main_res}
Following Algorithm~\ref{alg:HL_Zeno}, to achieve with high probability a global reconstruction precision on the Hamiltonian coefficients of order
\[
O\!\Bigl(n_p \sum_{i=1}^{n_p}\varepsilon_i\Bigr),
\]
it suffices to use a total number of copies
$O\!\bigl(\ln(n_p)/\epsilon_{\mathrm{QPT}}^{2}\bigr)$ and to perform
$O\!\bigl(1/\epsilon_{\mathrm{Z}}\bigr)$ kick repetitions. 
\end{main_res}
This ensures that the protocol exhibits no exponential dependence on the system size. Moreover, in the 1D case, increasing $k$ introduces only a constant multiplicative factor in the required number of copies (more details in Appendix \ref{sec:asymptotical}).

\begin{figure*}[t!]
\centering
    \includegraphics[width=0.9\linewidth]{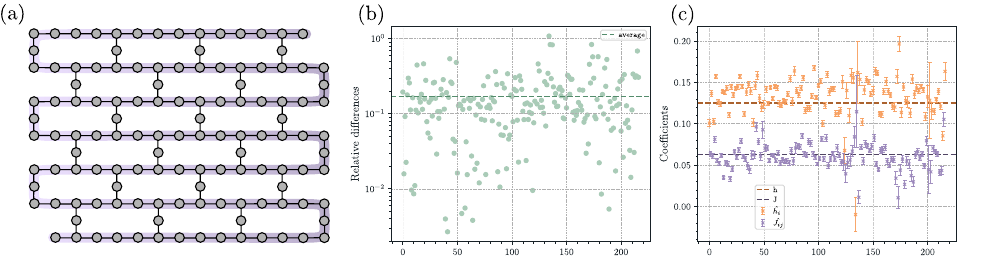}
    \caption{\justifying Experimental results. The experiment was performed with a total evolution time of $T=1$, $r=10$ kick repetition for the Zeno reshaping and $900$ shots for each setting ($36\times9\times3$) on the \texttt{ibm\_brisbane} device. (a). Selection of qubits used to perform the 109-qubit experiment. (b). Relative differences between real and reconstructed coefficients. (c). Reconstructed Hamiltonian coefficients: one-body (orange) and two-body (purple). Error bars denote one standard deviation obtained via Monte Carlo resampling \cite{Feigelson_Babu_2012} ($10^4$ instances).}
    \label{fig:exp_results}
\end{figure*}

\section{Numerical results}
\label{sec:numerics}

We numerically evaluate the performance of the protocol by reconstructing random 1D 2-local Hamiltonians for systems of sizes up to $N = 128$. Each Hamiltonian includes all possible single- and two-qubit Pauli terms, with coefficients drawn independently and uniformly from the interval $[-1,1]$. The time evolution duration is fixed to $T = 0.01$, chosen to satisfy the assumptions required for the convergence bounds discussed in \ref{coeff_bound}. Each data point corresponds to the average reconstruction error over 10 random Hamiltonians, and the quality of the reconstruction is defined as the Euclidean distance $\|\Delta c\|_2$ between the true and estimated coefficient vectors. The error bars in all plots represent the standard deviation across the set of random Hamiltonians used for averaging. We simulate the $N$-site chain as a finite‑size matrix‑product state and evolve it with the time‑evolving block decimation algorithm \cite{tebd}. 
All tensor‑network routines are provided by the TeNPy Python package \cite{tenpy}.

We evaluate the full protocol’s performance by analyzing the impact of both a finite number of kicks during reshaping and limited measurement statistics on the reconstruction error. As shown in Appendix~\ref{appendix:zenoscaling}, at \(r = 10\) kicks, the Zeno-induced error becomes negligible compared to the total error. Therefore, we fix the number of kicks at this value and focus on how sampling resources affect the reconstruction. Figure~\ref{fig:wide_plot}(a) shows the quality of reconstruction metric plotted against the total number of copies used.

As expected, the reconstruction quality improves as more resources are available for all system sizes. However, the overall norm is slightly larger for bigger systems, as more coefficients contribute to the vector. The inner panel in Figure~\ref{fig:wide_plot}(a) instead plots the average absolute error (difference between real and reconstructed coefficients) for a fixed number of copies and remains essentially constant as the system size grows. This is the anticipated behavior, as the protocol reconstructs each two-qubit patch independently and employs a fixed number of reshaping configurations for all $N>3$. 
To evaluate the reconstruction accuracy at the level of individual Hamiltonian terms, we consider the representative instance with $N = 9$, reconstructed using the full protocol with evolution time $T = 0.01$ and $10^{10}$ number of copies. The left panel of Fig.~\ref{fig:wide_plot}(b) shows the absolute differences $|\hat{c}_j - c_j|$ between the reconstructed and true coefficients for all single- and two-qubit Pauli terms. The dashed line indicates the average reconstruction error, which lies around $10^{-2}$, while many coefficients are recovered with even higher precision, reaching errors as low as $10^{-3}$–$10^{-4}$. The right panel compares the true and reconstructed coefficients directly, showing close agreement across the entire set. This detailed comparison highlights the precision of the learning procedure at the term-by-term level.

\section{Experimental results}
\label{sec:experimental}

We experimentally implement our Hamiltonian learning protocol on
an IBM Quantum device \cite{cloud_ibm}, accessed via the cloud and programmed using Qiskit \cite{javadiabhari2024quantumcomputingqiskit}. Specifically, we consider the $N$-qubit Hamiltonian
\begin{align}
    H = h \sum_{j=0}^{N-1} Z_j + J \sum_{j=0}^{N-2} X_j X_{j+1},  \label{eq:hamil_exp}
\end{align}
which models a linear chain of trapped ions with nearest-neighbor interactions. The dynamics are implemented using Trotterization \cite{trotter1959product, suzuki1976generalized}, where each Trotter step is immediately followed by a unitary kick. This discretization ensures that the number of kicks matches the number of Trotter steps. 

Our protocol is implemented on IBM's 127-qubit processor, employing dynamical decoupling to mitigate decoherence and randomized compiling to convert coherent errors into stochastic noise. This permits us to model the noisy evolution as a unitary channel subject to depolarizing noise \cite{Kim_2023}. To apply the learning procedure, the experimentally reconstructed Choi matrix is projected onto the closest unitary channel, thereby isolating the unitary component of the evolution and filtering out non-unitary effects. This post-processing step enables robust estimation of the Hamiltonian parameters despite hardware imperfections.

We performed the experiment on a chain of 109 qubits, applying the Zeno reshaping procedure with $r=10$ kicks and a total evolution time of $ T = 1 $. The Hamiltonian parameters were chosen as $ h = 1/8 $ and $ J = 1/16 $ to satisfy the conditions outlined in Section \ref{sec:bound}. To run the protocol, every setting of fiducial state, measurement operator, and reshaping configuration was executed with 900 shots on the hardware. The entire experiment was executed in 4 minutes and 24 seconds. Figure~\ref{fig:exp_results} (a) shows the specific selection of physical qubits used on the \texttt{ibm\_brisbane} processor to implement the chain. In panel Fig.~\ref{fig:exp_results} (b) we quantify the reconstruction error via the relative difference $\delta_j \;=\;\bigl|c_j - \hat c_j\bigr|/\lvert c_j\rvert$.  This normalized metric compensates for the inherently different scales of one‑ and two‑body terms.  The distribution of \(\delta_j\) shows that the protocol achieves an average relative error around $10\%$. Panel Fig.~\ref{fig:exp_results} (c) instead shows the reconstructed coefficients \(\{\hat c_j\}\) grouped by interaction order: one‑body terms \(h_i\) (orange) and two‑body couplings \(J_{ij}\) (purple). The dashed horizontal lines refer to the true parameter values. We can see that the experimentally estimated parameters are close to their theoretical values, having average error $|h-\hat{h}| \approx 0.021\pm0.005$ and $|J-\hat{J}| \approx 0.012\pm0.004$. Notice that these small errors with narrow error bars are only possible thanks to randomized compiling and error mitigation. Without them, error dominates, and the evolution according to the Hamiltonian $H$ of Eq.\thinspace\eqref{eq:hamil_exp} is lost. Therefore, our protocol validates the use of random compilation and error mitigation to improve results in quantum simulation. More details on the hardware implementation are provided in Appendix~\ref{appendix:hardware}.

\section{Discussion and Conclusion}\label{sec:discussion}

In this work, we present a protocol for efficient HL in current quantum hardware. It relies on a reshaping strategy based on a coherent implementation of the QZE to confine the dynamics to spatially disjoint local patches, whose Hamiltonian coefficients are learned in parallel via QPT. We establish performance guarantees for the reconstruction accuracy as a function of the number of copies and the number of Zeno kicks employed in the protocol. We validate its performance through numerical simulations of random, geometrically $k=2$ local Hamiltonians in 1D chains of up to 128 qubits. Our results show that the reconstruction quality remains consistent across different system sizes and that the average error of the protocol remains constant. To further support the feasibility of our approach, we implemented the protocol on IBM’s 127-qubit superconducting hardware, successfully learning a 109-qubit Hamiltonian. This demonstrates that our protocol is both scalable and suitable for large-scale state-of-the-art devices, which is especially important as the need to characterize increasingly large quantum systems continues to grow.

This suitability is ensured by the use of hardware-friendly operations. Since the reshaping used in our protocol relies solely on $Z$ rotations, the unitary kicks required to realize the Zeno effect can be efficiently implemented as virtual $Z$ gates.
These gates act instantaneously and do not introduce additional decoherence. Moreover, the use of local state preparation together with standard measurement schemes in process tomography, both avoiding the need for entangling operations, help minimize the overall reconstruction error. Collectively, these features make our protocol suitable for characterizing near-term quantum devices.

The scalability of our method arises from the reshaping of the dynamics induced by the Zeno effect, which effectively reduces the exponential complexity of the learning task \cite{Eisert_2020_cetification}. As a result, the required resources and the reconstruction accuracy depend just polynomially on the system size. This enables us to perform the quantum processing and classical postprocessing required by our learning scheme efficiently and implement the protocol on IBM quantum devices with up to hundreds of qubits. These experimental results highlight the potential of our technique as a benchmarking tool for recent large-scale digital quantum simulation experiments that already reach this regime \cite{Kim_2023, Chowdhury2024_sim2, Miessen2024_sim3, Cobos2025_sim4}. 

Previously, other Hamiltonian learning protocols have sought scalability by exploiting locality in different ways. Some of these approaches employ Hamiltonian reshaping through random gate sequences, combined with quantum-enhanced phase estimation, to achieve the theoretically optimal Heisenberg scaling \cite{Huang_2023_heisemberscaling}. While optimal, these methods require long coherent evolutions and frequent non-virtual gates, which increase exposure to noise and make them difficult to implement on current hardware. Alternative strategies exploit the locality guaranteed by the Lieb–Robinson bound \cite{Lieb1972, Poulin2010, Kuwahara2020}, restricting the learning to small subsystems within an effective light cone. This truncation introduces an error that grows exponentially with the evolution time, and limits the achievable precision over longer dynamics \cite{Wiebe2015, Mobus2023, StilckFrana2024}. In contrast, our reshaping technique confines the system’s dynamics through frequent virtual $Z$ gates, effectively suppressing unwanted couplings without increasing the total evolution time. This yields a more favorable scaling of the reconstruction error with the evolution duration compared to Lieb–Robinson–based approaches.

Other methods instead reconstruct the Hamiltonian from Gibbs or thermal states \cite{Bairey_2019_localmeasurements, Anshu_2021_gibbs, Haah_2024_gibbstates, Qi_2019_fromeigenstate}.  Although these states are informationally complete, their preparation on current devices is challenging, requiring either long thermalization times or complex ancilla-assisted circuits \cite{Wang2021, Holmes2022, Consiglio2024, Brunner2025}. Our approach avoids these limitations by relying only on simple product states and short coherent evolutions.

Further improvements of the protocol are possible along two main directions.
First, when targeting Hamiltonians containing higher-order $k$-body terms ($k>2$), such as those arising in quantum-chemistry models \cite{jordan1928uber}, the exponential scaling of standard QPT can be mitigated by adopting tensor-network process tomography \cite{Cramer2010, Ohliger2012Efficient, Lanyon2017Efficient, Torlai2023Process, Dang2022Process7Qubit}.
This method efficiently estimates multi-qubit processes that admit tensor-network representations with low bond dimension, reducing the number of required measurements from exponential to polynomial while preserving local state preparation and measurement.

Second, the sampling scaling of our protocol can be improved beyond the scaling of standard QPT.
By replacing the tomography step with the bootstrapped estimation algorithm of \cite{haah2023query}, one could, in principle, reach the Heisenberg limit, provided that controlled powers of the unitary evolution are accessible.

A natural direction for extending our approach is to consider Hamiltonians defined on more general geometries.
While this work focused on one-dimensional chains, the proposed reshaping technique can be straightforwardly adapted to higher-dimensional graphs.
For example, in square \cite{Google_2024} or hexagonal \cite{Hetenyi2024} qubit lattices, the layout can be partitioned into sets of non-overlapping linear chains, on which the reshaping is applied independently.
Since three configurations suffice to reconstruct all two-body interactions within a single chain, four distinct tilings of such chains are enough to cover an entire two-dimensional grid.
Consequently, the full lattice can be characterized with twelve reshaping configurations in total, regardless of its size, preserving the constant-overhead scaling demonstrated for the one-dimensional case.
For more general coupling graphs \cite{Hazra2021, Ramette2022, He2024, Wu2024, Lobe2024, Wang2024}, the same idea extends naturally through graph-coloring algorithms \cite{Ni2024}, which determine the minimal set of parallelizable configurations required for full Hamiltonian reconstruction.

Beyond static Hamiltonians, our protocol naturally generalizes to the estimation of time-dependent generators and open-system dynamics.
By omitting the unitary projection in the post-processing stage, the reconstructed Choi matrices provide direct access to the local Lindbladians \cite{Breuer_Petruccione_2007}, enabling the characterization of dissipative processes relevant for noisy devices \cite{Eisert_2020_cetification}.
Similarly, combining the proposed reshaping technique with time-resolved measurement schemes \cite{deClercq2016, Che2021, Siva2023_weak} could enable efficient learning of time-dependent two-body interactions.

Zeno-based reshaping can also be adapted to analog quantum simulators \cite{Arg_ello_Luengo_2019_qchemistry, Hangleiter2022-dq}, where the dynamics unfolds continuously under a fixed Hamiltonian rather than through discrete gate sequences. In this context, strong continuous measurements  \cite{Slichter_2016_contmeas} can effectively replace the unitary kicks, confining the evolution to selected subspaces and enabling local Hamiltonian estimation. Practical limitations mainly arise from measurement back-action and readout errors, which must be properly accounted for in realistic implementations \cite{Convy2022}.

\section*{Data and code availability statement}
All code for the simulations and IBM-hardware experiments, together with a step-by-step tutorial notebook, is available at \cite{git_tuto}; the complete simulation and experimental datasets are archived on Zenodo \cite{franceschetto_2025_17351462}.

\begin{acknowledgments} 
This work was supported by the Government of Spain (Severo Ochoa CEX2019-000910-S, FUNQIP, Quantum in Spain and European Union NextGenerationEU PRTR-C17.I1), Fundació Cellex, Fundació Mir-Puig, Generalitat de Catalunya (CERCA program), the ERC AdG CERQUTE, and the EU (PASQuanS2.1 101113690 and Quantera Veriqtas projects), and the AXA Chair in Quantum Information Science. G.F. and E.P. acknowledge support from a “la
Caixa” Foundation (ID 100010434) fellowship. The fellowship codes are LCF/BQ/DI23/11990070 and LCF/BQ/DI23/11990078. We acknowledge the use of IBM Quantum services for this work. The views expressed are those of the authors, and do not reflect the official policy or position of IBM or the IBM Quantum team.
\end{acknowledgments}
\bibliography{bib}

\appendix
\section{Proof of theorem}
\label{sec::theoremproof}
In this section, we present the proof for Theorem \ref{thm:1}. The main objective is to determine the individual contributions of the reshaping process and coefficient learning to the overall achieved precision. We begin by separating the error resulting from the unitary projection of the linear inverted channel and the error associated with the linear inverted estimation relative to the actual process
\begin{equation} 
    \| \hat{\varLambda}_{\text{U1}}-\varLambda\|_\diamond \le \| \hat{\varLambda}_{\text{U1}}-\hat{\varLambda}_{\text{LI}}\|_\diamond + \|\hat{\varLambda}_{\text{LI}}-\varLambda\|_\diamond.
\end{equation}
Knowing that we can compute the first element a posteriori, we focus on the latter
\begin{align}
    \|\hat{\varLambda}_{\text{LI}}-\varLambda\|_\diamond \le \underbrace{\|\hat{\varLambda}_{\text{LI}}-\varLambda_{\text{Z}}\|_\diamond}_{(1)}+ \underbrace{\|\varLambda_{\text{Z}}-\varLambda\|_\diamond}_{(2)},
\end{align}
where $(1)$ can be bounded from the results presented in \ref{sub:coeff_learning}. For $(2)$, we consider that
\begin{align}
    \|\varLambda_{\text{Z}}-\varLambda\|_\diamond&\le2\|\operatorname{Tr_S(}V_r(T)-U_\text{Z}(T))\|_\text{op}\\
    &\le 2\underbrace{\|V_r(T)-U_\text{Z}(T))\|_\text{op}}_{(3)}
\end{align}
where $S$ is the complement of the target subsystem that we are learning and finally (3) can be bounded from \ref{sub:ham_reshaping}.

\section{Proof of corollary}\label{sec::corproof}
Beginning with Theorem \ref{thm:1}, it is possible to constrain the precision achieved on the coefficient vector, thereby providing a demonstration for Corollary \ref{coeff_bound}. By the definition of the diamond norm
\begin{equation}
    \|\hat{\varLambda}_\text{U1}-\varLambda\|_\diamond \ge \|U_\text{U1}-U_\text{patch}\|_\text{op}.
\end{equation}
Then, expressing the unitaries in terms of their Hamiltonians
\begin{equation}
    \|U_\text{U1}-U_\text{patch}\|_\text{op}=\|e^{-i\hat{\bar{H}}_\text{patch}}-e^{-i\bar{H}_\text{patch}}\|_\text{op},
\end{equation}
and, under the corollary conditions, we can apply Lemma 3.2 (b) of \cite{haah2023query}
\begin{equation}
    \|e^{-i\hat{\bar{H}}_\text{patch}T}-e^{-i\bar{H}_\text{patch}T}\|_\text{op}\ge \frac{T}{\pi}\|\hat{\bar{H}}_\text{patch}-\bar{H}_\text{patch}\|_\text{op},
\end{equation}
where 
\begin{equation}
   \bar{H}_\text{patch} = \sum_j c_j h_j,\quad \hat{\bar{H}}_\text{patch} = \sum_j \hat{c}_j h_j,\quad \Delta c = c - \hat{c}. 
\end{equation}
Then, applying the above definitions
\begin{equation}
    \|\hat{\bar{H}}_\text{patch}-\bar{H}_\text{patch}\|_\text{op} = \frac{1}{2^{n/2}}\|\sum_j \Delta c_j h_j\|_F,
\end{equation}
where $n$ is the number of qubit of the target subsystem and $\|\cdot\|_F$ the Frobenius norm.
By the definition of the Frobenius norm
\begin{align}
    \|\sum_j \Delta c_j h_j\|_F &= \sqrt{\operatorname{Tr\left(\left(\sum_j\Delta c_j h_j\right)^\dagger\left(\sum_j\Delta c_j h_j\right)\right)}}\\
    &=\sqrt{\left(2^n\sum_j\Delta c_j^2\right)}= 2^{n/2}\|\Delta c \|_2.
\end{align}
Finally
\begin{equation}
    \|\Delta c \|_2\le \frac{\pi}{T}\|\hat{\varLambda}_\text{U1}-\varLambda\|_\diamond ,
\end{equation}
that allow us to use the results from Theorem \ref{thm:1}.
\section{Discussion on the overall protocol asymptotic scaling}\label{sec:asymptotical}
This section analyzes the asymptotic scaling of the full protocol to justify the statements made in the main text. Let $\|\Delta c\|_2$ denote the norm of the difference between the reconstructed and true coefficients of the global Hamiltonian, and $\|\Delta c_{p_i}\|_2$ the corresponding norm restricted to the $i$-th patch. The total number of patches reconstructed locally is denoted by $n_p$, obtained through $n_c$ reshaping configurations (with $n_c = 3$ in the example presented in the main text).
Notice that the norm of the differences of the full set of coefficients can be upper bounded by $\|\Delta c \|_2\leq n_c \sum_{i=1}^{n_p} \|\Delta c_{p_i}\|_2$\footnote{This can be derived by looking at: 1. how to reconstruct the full set of coefficients from the local ones, essentially how you have to correct the errors appearing due to the reshaping side-effects and 2. how the difference vector of the full set of coefficients can be expressed in term of the difference vectors of the coefficients on the patches.}. Then, reminding that $\varepsilon_i = \frac{\pi}{T}\,\Bigl(\|\hat{\varLambda}_{\mathrm{U1}} - \hat{\varLambda}_{\mathrm{LI}}\|_\diamond + \epsilon \Bigr) $, we have
\begin{align}
    \Pr\Bigl[\| \Delta c \|_2 \geq  \sum_{i=1}^{n_p} \varepsilon_i  \Bigr] & \leq \Pr\Bigl[n_c \sum_{i=1}^{n_p} \Vert \Delta c_{p_i} \Vert_2 \geq \sum_{i=1}^{n_p} \varepsilon_i  \Bigr] \nonumber \\
    & \leq \sum_{i=1}^{n_p} \Pr\Bigl[ \Vert \Delta c_{p_i} \Vert_2 \geq \varepsilon_i/n_c   \Bigr]. 
\end{align}
The first inequality follows from the triangle inequality, and the second from the union bound. To guarantee that the error probability across the entire protocol does not exceed $\delta$, we require each term in the final sum to be bounded by $\delta/n_p$. Then, for this to be true, we need
    \[
        N_{\mathrm{copies}} \;\ge\; \frac{n_c^2 C_{\mathrm{QPT}}(n)}{\epsilon_{\mathrm{QPT}}^2}\,\ln\!\Bigl(n_p \frac{2^{4n}}{\delta}\Bigr). 
    \]
and 
\[
        r \;\ge\; \,\frac{2n_cC_{Z}}{\epsilon_{\rm{Z}}}.
    \]
\section{Scaling of Zeno error}
\label{appendix:zenoscaling}
\begin{figure}[t!]
    \centering
    \includegraphics[width=\linewidth]{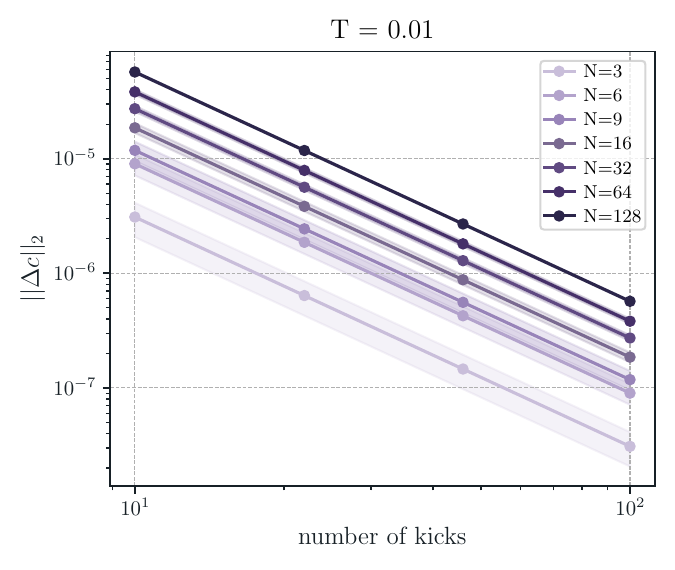}
    \caption{\justifying Euclidean norm of the difference between reconstructed and true Hamiltonian coefficients as a function of the number of kicks, for system sizes $N=3,6,9$ and fixed total evolution time $T=0.01$. The reconstruction is performed under ideal conditions with exact tomography and noiseless measurements, so the plotted error reflects only the contribution from imperfect Zeno confinement.}
    \label{fig:N_vs_proj}
\end{figure}
Figure~\ref{fig:N_vs_proj} shows the dependence of the reconstruction accuracy on the number of unitary kick rounds used. To isolate the effect of imperfect Hamiltonian reshaping, these simulations are carried out under idealized conditions, assuming noiseless measurements and exact QPT. In this setting, the reconstruction error reflects only deviations introduced by approximating the ideal Zeno dynamics with a finite number of unitary kicks. As expected, increasing the number of kicks improves the quality of the reconstruction across all system sizes. The minor differences observed across values of $N > 3$ arise from finite-size fluctuations and the increasing number of terms included in the comparison, rather than any intrinsic scaling limitation. With only tens of kick rounds, the quality of the reconstructions reaches a level where the associated error becomes negligible compared to the dominant contribution from measurement shot noise of QPT. This means that further increasing the number of kicks offers marginal returns, and using around 10 kicks is sufficient to ensure that Hamiltonian reshaping no longer limits the overall reconstruction quality.
\begin{figure*}
    \centering
    \includegraphics[width=\linewidth]{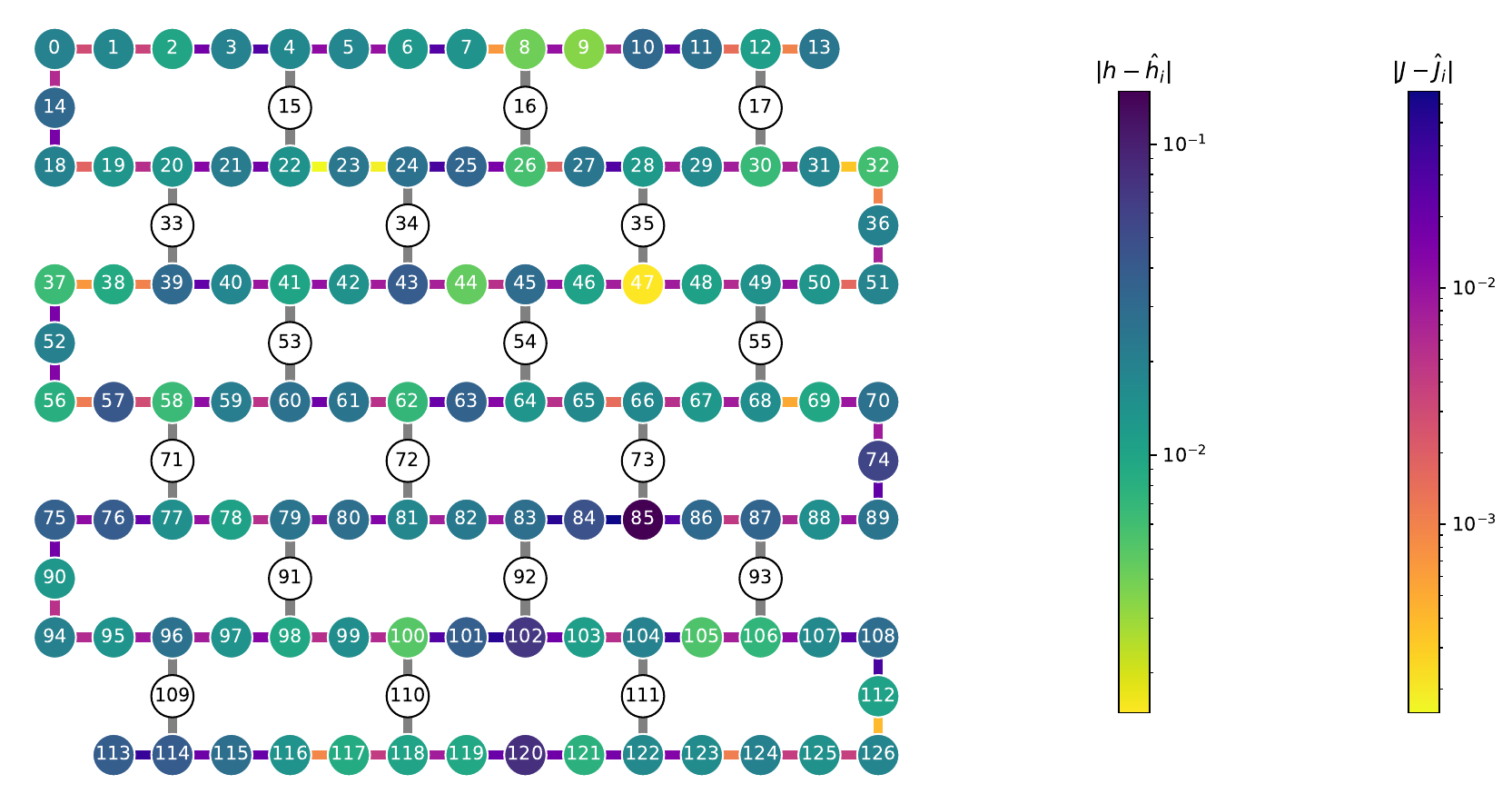}
    \caption{\justifying Experimental results of the reconstruction of a 109-qubit Hamiltonian. We represent the errors in the reconstructed coefficients through color maps on the hardware geometry. The circles stand for the one-body coefficients, while the connections represent the two-body couplings.}
    \label{fig:exp}
\end{figure*}

\section{Hardware implementation details}
\label{appendix:hardware}
The experiment was implemented on the 127-qubit \texttt{ibm\_brisbane} device, based on the Eagle R3 architecture. This device has a median single-qubit gate error of $2.256 \times 10^{-4}$ (X and SX), a two-qubit gate (ECR) error of $6.811 \times 10^{-3}$, and a readout error of $1.685 \times 10^{-2}$. The median qubit relaxation time is $T_1 = 219.48\,\mu$s. RZ gates are virtual, implementing the unitary kick with negligible error.
Due to the inherent noise in current quantum hardware, the implemented evolution deviates from unitarity, which violates the assumptions of our protocol. To mitigate this, we employ error suppression techniques to recover approximate unitary dynamics. First, digital dynamical decoupling is used to counteract decoherence by isolating qubits from the environment. Second, randomized compiling is implemented to transform coherent errors into stochastic noise, thereby making the effective dynamics better modeled by a depolarizing channel \cite{Kim_2023}. Under this approximation, the process can be modeled by a noisy Choi matrix
\begin{align}
    \Upsilon = (1 - \lambda) \Upsilon_U + \lambda \Upsilon_{\text{noise}},
\end{align}
where $\Upsilon_U$ is the ideal unitary part, $\Upsilon_{\text{noise}}$ captures non-unitary contributions, and $\lambda$ quantifies the noise strength. To eliminate the non-unitary evolution $\Upsilon_{\text{noise}}$, we adopt a post-processing strategy. We search for the closest unitary channel to the experimentally estimated Choi matrix $\hat\Upsilon$ by solving the optimization problem
\begin{align}
    \hat{U} = \arg\max_{U \in \mathcal{U}(d)} \bra{U} \hat{\Upsilon} \ket{U},
\end{align}
where $\ket{U}$ denotes the vectorization of the unitary $U$. The noise strength is then estimated as
\begin{align}
    \hat{\lambda} = 1 - \frac{1}{d} \bra{\hat{U}} \hat{\Upsilon} \ket{\hat{U}}.
\end{align}
This approach enables us to approximately recover the ideal unitary evolution, allowing for reliable Hamiltonian learning despite hardware imperfections. We present the results from the main text, accompanied by a color map of the device geometry in Figure~\ref{fig:exp}. The median of the estimated parameters is $\hat h \approx 0.133\mp 0.005$ and $\hat J \approx 0.060\mp0.004$. The median of the error for each parameter is $|h-\hat{h}| \approx 0.021\pm0.005$ and $|J-\hat{J}| \approx 0.012\pm0.004$. We estimate a median noise strength of $\lambda \approx 0.556\pm0.004$. Error bars denote one standard deviation obtained via $10^4$ instances of Monte Carlo resampling \cite{Feigelson_Babu_2012}. Plotting the errors as a color map on the device's layout provides intuition about which regions of the device are more error-prone.

\end{document}